# Emergent Gauge Fields

**Peter G.O. Freund**

***Enrico Fermi Institute and Department of Physics, University of Chicago, Chicago, IL 60637***

**Abstract:** Erik Verlinde's proposal of the emergence of the gravitational force as an entropic force is extended to abelian and non-abelian gauge fields and to matter fields. This suggests a picture with no fundamental forces or forms of matter whatsoever.

In Maldacena's famous AdS/CFT duality [1] a string theory which necessarily includes gravity has as its dual counterpart a field theory *without* gravity. Gravity, rather than playing the role of a fundamental force, then becomes something that can be removed by simply switching to an equivalent dual picture. This, along with the holographic approach [2] it leads to, suggests that gravity should be an emergent force.

In a remarkable paper [3] E. Verlinde, building on earlier work by Jacobson [4] and Padmanabhan [5] has proposed a specific mechanism for the *emergence* of gravity *and* of space-time in a theory containing matter: gravity is an entropic force. This way he has been able to reproduce both Newton's nonrelativistic gravitational force law and Einstein's generally covariant equations. At face value, this would distinguish gravity form the electro-weak and color forces, which would maintain their fundamental-force status.

Here we wish to point out that the entropic force picture can lead also to the emergence of the electro-weak and color forces, and of their Yang-Mills gauge equations.

This observation can be justified in a very simple manner. Verlinde's argument is valid in any number of space-time dimensions [3], though, as noted in ref. [5], in higher dimensions the generic entropic argument leads to the more general Lanczos-Lovelock theories of which Einstein's theory is a special case. In any case, one can then see emerging space-times of any number *D* of dimensions along with their gravities. In particular, a certain number, say *D-d,* of the emerging dimensions may be compact. As far as the *d* non-compact dimensions are concerned, it will appear that the field equations of the higher *D*-dimensional theory (which includes both the non-compact and the compact dimensions) reduces à la Kaluza-Klein [6], up to some scalar fields, to a coupled gravity-Yang-Mills type system, with the Yang-Mills fields as components of *D*-dimensional gravity. Being components of emergent *D*-dimensional entropic gravity, these Yang-Mills gauge fields are therefore also emergent rather than "fundamental."

It should be stressed that the emergent nature of these gauge fields stems from the fact that in a higher dimension they are components of gravity. This is the important ingredient of our observation, since an attempt to find a *four-dimensional emergent* gauge field *directly* in four dimensions in the absence of any additional dimensions, runs into some difficulties [7] because the appearance of charges of opposite sign for particles and their antiparticles seems to lead to negative temperatures.

To see how this difficulty is avoided in a theory with extra space dimensions let us consider a typical and very familiar example: the reduction of Kaluza and Klein's original 5-dimensional gravity to 4 dimensions. The 5-dimensional gravitational field is even under charge conjugation and as such the entropic argument for its emergence works without further ado. Yet, as known already to Klein, upon reduction to 4 dimensions, the higher harmonics in the periodic fifth coordinate correspond to "electrically" charged fields of *both s*igns, which in a direct 4-dimensional argument would cause the just mentioned difficulty. Indeed in the 4-dimensional; reduced theory we encounter fields which are even, as well as fields which are odd under the 4-dimensional charge conjugation. This is related to the old observation [8] that the charge-sign-reversing and seemingly non-geometric charge conjugation operation in 4 dimensions is obtained by dimensional reduction from a harmless geometric reflection of the extra "internal" fifth space dimension. For more general *D*-dimensional theories in which all fields are even under charge conjugation to start with, the entropic force picture is obtained in the full *D* dimensional space, and the resulting field theory is *then* dimensionally reduced to a theory in $d<D$ dimensions, which contains also charged particles and their *oppositely charged* antiparticles! Along with bosonic fields even under $d$-dimensional charge conjugation, the dimensionally reduced theory also contains gauge fields odd under $d$-dimensional charge conjugation. In particular, $D=11$ supergravity is of this type: all fields that appear in this $11$-dimensional theory are even under charge-conjugation. $D=11$ is suggested by the $M$-theory that string theory leads to. The case $d=4$ may even be of phenomenological interest. It should also be mentioned that for $D$-dimensional gravity of the Lanczos-Lovelock type ($D>4$), the Yang-Mills dynamics acquires extra, still gauge-invariant but higher order, interaction terms.

It would seem then that in such a generalized picture we encounter no fundamental fields whatsoever, all forces are emergent! In fact, once one accommodates enough supersymmetry in these considerations, further gauge fields and even all matter fields might be emergent. This is just the opposite of the old approach based on the idea of a set of fundamental forces and of fundamental forms of matter.

It is as if assuming certain forces and forms of matter to be fundamental is tantamount (in the sense of an effective theory) to assuming that there are no fundamental forces or forms of matter whatsoever, and everything is emergent. This latter picture in which nothing is fundamental is reminiscent of Chew's bootstrap approach [9], the original breeding ground of string theory. Could it be that after all its mathematically and physically exquisite developments, string theory has returned to its birthplace?

I wish to thank Erik Verlinde for very helpful correspondence from which it is clear that he independently has also arrived at the conclusion that not only gravity, but all gauge fields should be emergent. My thanks are also due to Jeff Harvey and Reinhard Oehme for useful comments.